\documentclass[sigconf,natbib=true,anonymous=false]{acmart}

\usepackage{enumitem}
\usepackage{booktabs}
\usepackage{graphicx}
\usepackage{array}
\usepackage{hhline}
\usepackage{makecell}
\usepackage{ulem}
\usepackage{float}
\usepackage{newfloat}

\usepackage{multirow}
\usepackage{cuted}   % 提供 strip 环境

\usepackage[linesnumbered,ruled]{algorithm2e}
\usepackage{algorithmic}
\usepackage{amsmath}
\usepackage{amssymb}
\SetKwInput{KwRequire}{Require} % 定义Require关键字
\SetKwInput{KwResult}{Result}   % 定义Result关键字

\AtBeginDocument{%
  }

\copyrightyear{2026}
\acmYear{2026}
\setcopyright{cc}
\setcctype{by}
\acmConference[SIGIR '26]{Proceedings of the 49th International ACM SIGIR Conference on Research and Development in Information Retrieval}{July 20--24, 2026}{Melbourne, VIC, Australia}
\acmBooktitle{Proceedings of the 49th International ACM SIGIR Conference on Research and Development in Information Retrieval (SIGIR '26), July 20--24, 2026, Melbourne, VIC, Australia}
\acmDOI{10.1145/3805712.3809566}
\acmISBN{979-8-4007-2599-9/2026/07}

\begin{document}

\title{Generative Bid Shading in Real-Time Bidding Advertising}


\author{Yinqiu Huang}
\affiliation{%
\institution{Meituan}
   \city{Chengdu}
  \country{China}
  }
\email{huangyinqiu@meituan.com}

\author{Hao Ma}
\affiliation{%
  \institution{Chongqing University}
   \city{Chongqing}
  \country{China}
}
\email{ma_hao@cqu.edu.cn}

\author{Wenshuai Chen}
\affiliation{%
\institution{Meituan}
   \city{Chengdu}
  \country{China}
  }
\email{chenwenshuai@meituan.com}

\author{Zongwei Wang}
\authornote{Corresponding author.}
\affiliation{%
\institution{Chongqing University}
   \city{Chongqing}
  \country{China}
  }
\email{zongwei@cqu.edu.cn}

\author{Shuli Wang}
\affiliation{%
  \institution{Meituan}
  \city{Chengdu}
  \country{China}
}
\email{wangshuli03@meituan.com}

\author{Yongqiang Zhang}
\affiliation{%
\institution{Meituan}
   \city{Chengdu}
  \country{China}
  }
\email{zhangyongqiang08@meituan.com}

\author{Xue Wei}
\affiliation{%
\institution{Meituan}
   \city{Chengdu}
  \country{China}
  }
\email{weixue06@meituan.com}

\author{Yinhua Zhu}
\affiliation{%
\institution{Meituan}
   \city{Chengdu}
  \country{China}
  }
\email{zhuyinhua@meituan.com}

\author{Haitao Wang}
\affiliation{%
\institution{Meituan}
   \city{Chengdu}
  \country{China}
  }
\email{wanghaitao13@meituan.com}

\author{Xingxing Wang}
\affiliation{%
\institution{Meituan}
   \city{Beijing}
  \country{China}
  }
\email{wangxingxing04@meituan.com}

\renewcommand{\shortauthors}{Yinqiu Huang et al.}

\begin{abstract}
Bid shading plays a crucial role in Real-Time Bidding (RTB)  by adaptively adjusting the bid to avoid advertisers overspending.  Existing mainstream two-stage methods, which first model bid landscapes and then optimize surplus using operations research techniques, are constrained by unimodal assumptions that fail to adapt for non-convex surplus curves and are vulnerable to cascading errors in sequential workflows.      Additionally, existing discretization models of continuous values ignore the dependence between discrete intervals, reducing the model’s error correction ability, while sample selection bias in bidding scenarios presents further challenges for prediction.  To address these issues, this paper introduces Generative Bid Shading (GBS), which comprises two primary components: 1) an end-to-end generative model that utilizes an autoregressive approach to generate shading ratios by stepwise residuals, capturing complex value dependencies without relying on predefined priors;      and 2) a reward preference alignment system, which incorporates a channel-aware hierarchical dynamic network (CHNet) as the reward model to extract fine-grained features, along with modules for surplus optimization and exploration utility reward alignment, ultimately optimizing both short-term and long-term surplus using group relative policy optimization (GRPO).
Extensive experiments on both offline and online A/B tests validate GBS’s effectiveness.   Moreover, GBS has been deployed on the Meituan DSP platform, serving billions of bid requests daily.
\end{abstract}

\begin{CCSXML}
<ccs2012>
   <concept>
       <concept_id>10002951.10003227.10003447</concept_id>
       <concept_desc>Information systems~Computational advertising</concept_desc>
       <concept_significance>500</concept_significance>
       </concept>
 </ccs2012>
\end{CCSXML}

\ccsdesc[500]{Information systems~Computational advertising}


\keywords{Bid Shading, Generative Model, Auto-bidding}



\maketitle

\section{Introduction}
% Online advertising is now a primary way for companies to attract users and build their brands. Real-Time Bidding (RTB) is the key system in programmatic advertising, allowing auctions in milliseconds. When a user visits a website, the Supply-Side Platform (SSP) sends an ad request to the Ad Exchange (ADX), where Demand-Side Platforms (DSPs) bid based on request information. The highest bidder wins the chance to show their ad. The auction way has changed a lot due to transparency issues in auctions and compatibility challenges with emerging technologies \cite{r2, r3, r4}. Most major platforms, following Google \cite{r1}, have switched from Second-Price Auctions (SPA) to First-Price Auctions (FPA).   In SPA, advertisers usually bid through their actual value, but in FPA or non-ideal SPA settings, this can cause them to pay much more than necessary. For example, if an advertiser bids \$5 while the second-highest bid is only \$2, they would overpay up to \$3. By dynamically adjusting bids toward the minimum winning price, bid shading \cite{crespi2005multinomial} significantly optimizes cost structures and directly improves resource allocation efficiency in the trillion-dollar digital advertising market.
Online advertising has emerged as a cornerstone of the digital economy, with Real-Time Bidding (RTB) serving as the fundamental infrastructure for programmatic media buying. In this ecosystem, Demand-Side Platforms (DSPs) compete in auctions to purchase impressions on behalf of advertisers. Under the conventional RTB paradigm, these high-frequency auctions were typically implemented using a second-price rule, under which the highest bidder won the impression but paid a settlement price equal to the second-highest bid. This effective mechanism can incentivize truthful bidding by aligning bidders’ dominant strategy with submitting offers that reflect their underlying valuation of each impression, thereby mitigating the risk of systematic overpayment relative to prevailing competitive conditions.
% Online advertising has emerged as a cornerstone of the digital economy, with Real-Time Bidding (RTB) serving as the fundamental infrastructure for programmatic media buying. Conventionally, these instantaneous auctions operated under a pricing protocol in which the winning bidder—despite submitting the highest offer—was charged a settlement price equivalent to the second-highest bid. This second-price mechanism was designed to encourage truthful bidding, theoretically allowing advertisers to submit bids reflecting their true valuation of an impression without the immediate risk of overpaying relative to the market competition.

However, the practical application of second-price auctions has been increasingly challenged by a lack of transparency regarding settlement prices and an inability to adapt to complex, emerging auction environments \cite{r1, r2, r3, r4}. Recently, the industry has shifted to First-Price Auctions (FPA), where the winner pays their own bid, implying that bidding one’s full valuation generally reduces surplus and results in systematic overpayment. In an idealized setting where the market-clearing (minimum winning) price for a given bid request was observable, an advertiser could bid exactly that price to win the impression at the lowest possible cost. However, because this clearing price is latent and should be inferred under substantial uncertainty, advertisers instead rely on bid shading \cite{crespi2005multinomial, huang2024second}, which strategically discounts the submitted bid below the raw valuation. By calibrating this discount to trade off win probability against surplus preservation, bid shading can substantially reduce acquisition costs and improve budget efficiency, thereby enhancing resource allocation in the trillion-dollar digital advertising market.

\begin{figure}[!t]
\centering
\includegraphics[scale=0.32]{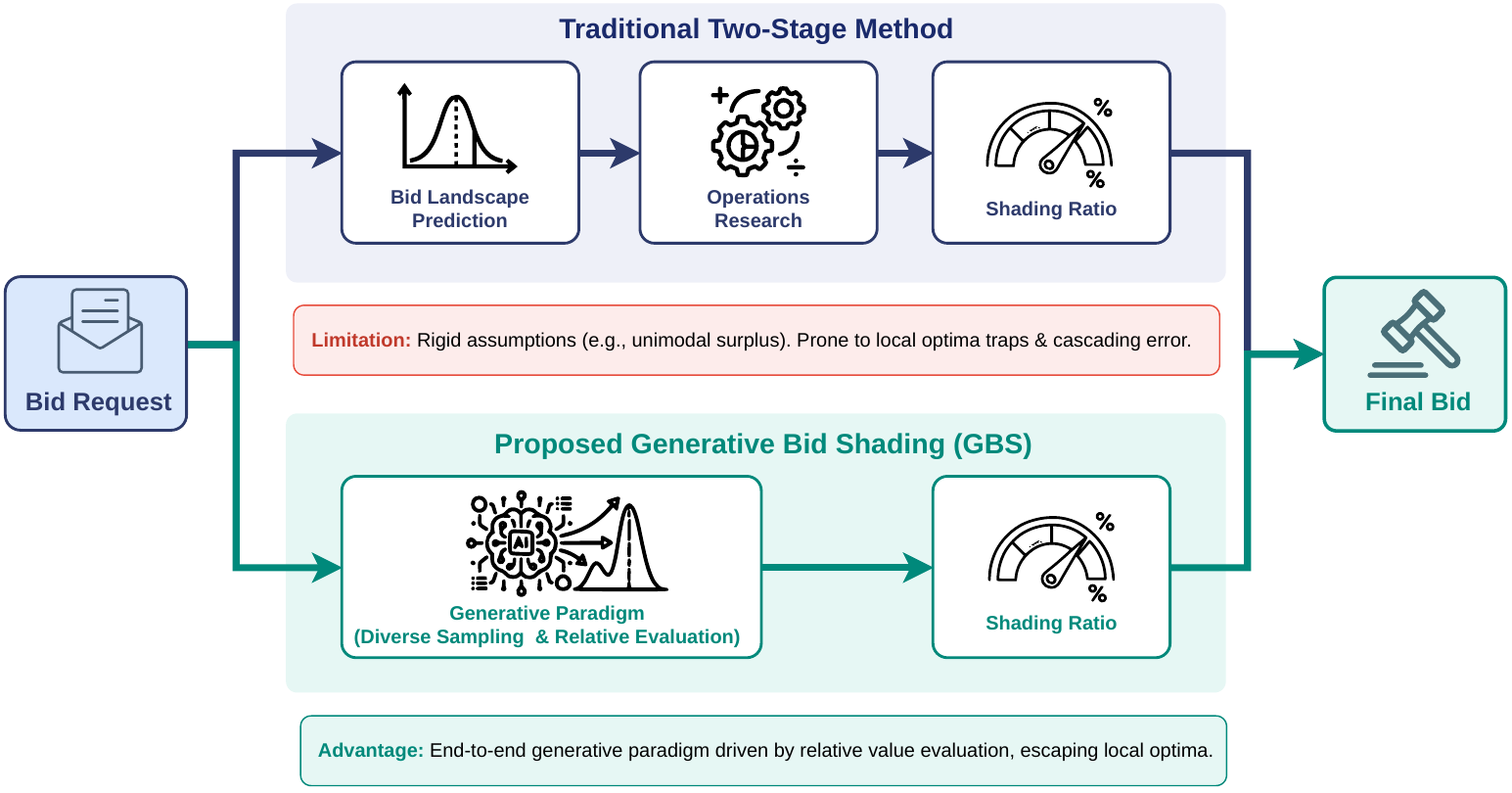}
\caption{Overview of bid shading process.}
\label{Fig1}
% \vspace{-0.5cm}
\end{figure}

% Bid shading is a censored problem \cite{huang2024second} because of the inaccessibility of the minimum winning price. Some methods assume an uncensored scenario in which the ADX will return the minimum winning price regardless of the auction outcome, allowing them to fit an optimal shading ratio using linear regression. However, this assumption is unrealistic in real-world DSP environments: in FPA, the minimum winning price is unobservable regardless of whether the bid wins or loses, resulting in no ground truth for bid shading. 

\begin{figure}[!t]
\centering
\includegraphics[scale=0.45]{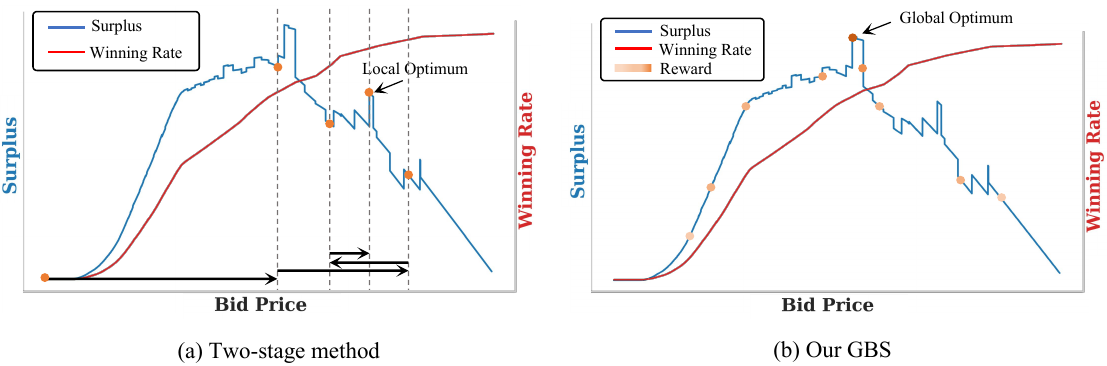}
\caption{The winning rate and surplus curves derived from  Meituan's non-ideal second-price auctions data.}
\label{Fig2}
% \vspace{-0.5cm}
\end{figure}

The prevailing approach to bid shading typically adopts a two-stage pipeline (Fig. \ref{Fig1}): it first estimates the bid landscape, and then uses an operations-research (OR) module to maximize surplus via search algorithms (e.g., bisection search \cite{pan2020bid}, golden section search \cite{zhou2021efficient}). However, this decoupled design is inherently brittle. Its rigidity leads to a key limitation: the final bid can be suboptimal because the optimizer relies on simplifying assumptions and is sensitive to estimation errors. 
Many OR solvers assume that the surplus as a function of the bid is unimodal \cite{pan2020bid, zhou2021efficient}, so that simple search procedures can reliably locate the optimum. In practice, the surplus curve in RTB is often non-convex and may contain multiple local peaks, causing search algorithms to converge to a local optimum rather than the global one. As illustrated in Fig. \ref{Fig2}, the surplus curve initially increases and then decreases, but does not strictly adhere to a standard unimodal pattern. In principle, one could evaluate all candidate bids online to find the best bid, but exhaustive evaluation is computationally infeasible under millisecond-level latency constraints. Moreover, because the workflow is sequential, errors in the first-stage estimation are directly carried into the second-stage optimization and can be further magnified, preventing the system from achieving practical optimality within strict latency budgets.

To address these issues, we explore a generative paradigm that mitigates local optima by combining diverse candidate generation with relative evaluation. In this paradigm, we sample multiple bid candidates from a generative model and evaluate them jointly, which approximates a broader search than single-point optimization. We further compute relative advantages among candidates, so learning is driven by comparisons rather than fragile absolute targets, reducing the chance of being misled by local extrema. As illustrated in Fig. \ref{Fig2}, conventional search methods frequently stagnate in local optima when facing non-convex objective functions, while GBS adopts a broader view of the generative space, allowing it to effectively converge to the global optimum. However, implementing this idea is non-trivial with existing generative architectures. Methods such as CVAE \cite{sohn2015learning} often rely on rigid predefined priors, while discretization-based approaches \cite{sun2024cread, lin2023tree} fail to leverage conditional dependencies between discrete intervals, as they treat these intervals as isolated labels without mutual interdependence.
Moreover, training data collected from historical winning bids is inherently biased and covers only a limited portion of the counterfactual action space, further restricting effective learning.

% break the natural ordering of bids and weaken fine-grained correction. 

% A recent study \cite{gong2023mebs} has attempted to generate the shading ratio in an end-to-end manner.   However, their use of negative log-surplus as the loss function in supervised learning encounters the same limitations as two-stage methods when faced with non-convex surplus curves: convergence to local optima.   Furthermore, their simple discriminative models struggle to meet the high-precision requirements of bidding scenarios.   Traditional generative algorithms, such as conditional variational autoencoders (CVAE \cite{sohn2015learning}), are constrained by predefined prior distributions (e.g., Gaussian), which limits their performance ceiling in shading ratio generation. To overcome distributional assumptions and mitigate outlier interference, existing direct scalar-output methods often discretize continuous values \cite{sun2024cread, lin2023tree}.   However, this approach fails to leverage conditional dependencies between discrete intervals;  discretization is only reflected in label definitions without mutual interdependence (e.g., 0.3 is closer to 0.4 than 0.2), and the lack of error-correction mechanisms leads to suboptimal results. Additionally, DSPs optimize holistic revenue during bidding, causing model performance to be heavily influenced by bid price distributions.   This introduces significant data selection bias, as the training data fails to represent the real action space of bids.

Building upon the above insights, we incorporate autoregressive modeling \cite{ma2024generative} and reinforcement-based alignment \cite{deng2025onerec} to propose a novel end-to-end framework, Generative Bid Shading (GBS). This approach leverages a global perspective to address the limitations of traditional two-stage methods, utilizing a generative architecture to optimize bid shading. GBS comprises two main components: 1) An autoregressive generative model: To address the limitations of rigid priors and the neglect of dependencies between discrete intervals, we propose an autoregressive model. By decomposing shading ratios into sequences of discrete tokens, our method treats generation as a conditional sequential process. This design enables the model to effectively capture high-order dependencies between token values, facilitating precise approximation through residual refinement. To support this, we employ teacher forcing and gumbel-softmax to ensure stable convergence and end-to-end differentiability. 2) A reward preference alignment system: We propose a reinforcement learning-based preference alignment system anchored by the Channel-aware Hierarchical Dynamic Network (CHNet), a high-precision reward model. Policy optimization is driven by two distinct alignment mechanisms: a surplus optimization alignment that maximizes immediate returns by curbing overbidding on obtainable requests, and an exploration utility alignment that integrates strategic bid exploration to counteract sample selection bias.    Furthermore, we employ Group Relative Policy Optimization (GRPO) to guarantee efficient and stable policy updates throughout the training process. Crucially, this design inherently circumvents the issues of local sub-optimality and cascading errors that typically plague two-stage approaches.

In summary, our contributions are as follows:

% \begin{itemize}[leftmargin=12pt]
\begin{itemize}
\item{
To our best knowledge, we are the first to formally identify the inherent limitations of the conventional two-stage bid shading paradigm.  We demonstrate that the decoupling of estimation and optimization leads to cascading errors, while the reliance on unimodal assumptions often results in local optimality within non-convex surplus landscapes, thereby fundamentally constraining the performance upper bound of existing methods.
% We identify the suboptimality in two-stage bid shading methods and propose a generative framework that autoregressively generates shading ratios with error correction capabilities from a global perspective.
}
\item{
% We design a reward model to capture granular features within the dynamic bidding landscape. Based on this, we introduce two reward preference alignment modules: one maximizes immediate surplus by avoiding overbidding; the other mitigates data selection bias via strategic bid exploration. This paradigm enhances generalization across diverse business scenarios and extensibility.
We propose a novel end-to-end generative paradigm designed to transcend traditional limitations.   This framework synergistically integrates a novel autoregressive generative model, a high-precision reward model (CHNet), and dual preference alignment mechanisms, which simultaneously maximize immediate surplus and mitigate data selection bias, achieving superior scalability and generalization across diverse business scenarios.
}
\item{
Through extensive experiments on both offline and online A/B tests, we show that our GBS significantly outperforms the state-of-the-art methods. GBS has been deployed in Meituan DSP and serves billions of bid requests daily. To the best of our knowledge, this work represents the first implementation of generative bid shading in large-scale online systems.
}
\end{itemize}

\section{Related Work}
\textbf{Bid Landscape Forecasting}.
Bid landscape forecasting is a crucial component of RTB advertising systems, primarily aiding in forecasting the winning price distribution of bid requests. Early research often assumed that winning price followed specific probability distributions, such as Gaussian \cite{wu2015predicting}, Log-Normal \cite{cui2011bid}, Gamma \cite{chapelle2015offline,zhu2017gamma}, or other parametric distributions \cite{ren2017bidding, ren2016user}. To overcome the limitations of these predefined distributional assumptions, subsequent studies introduced distribution-free models \cite{ren2019deep, huang2024second, li2022arbitrary}. However, existing methods generally overlook the modeling of fine-grained features, so we propose a channel-aware hierarchical dynamic network to achieve fine-grained modeling as a foundation for the reward system.

\textbf{Bid Shading}.
Bid shading helps advertisers avoid overpaying. Existing methods fall into three categories: 1) Direct regression of the shading ratio \cite{gligorijevic2020bid}, which predicts the optimal ratio but unrealistically assumes access to all winning prices; 2) Two-stage ML + OR optimization \cite{pan2020bid, zhou2021efficient, si2023optimal}, where machine learning estimates bid landscape and operations research maximizes surplus; and 3) Non-parametric estimation \cite{zhang2021meow}, which uses dynamic binning without parametric models, but still needs optimal shading labels and cannot fully capture individual differences.

\textbf{Generative Model and Preference Alignment}.
With the rapid development of large language models (LLMs), generative models are increasingly used in areas such as generative recommendation \cite{deng2025onerec} and search \cite{guo2025onesug}. Preference learning, including Reinforcement Learning from Human Feedback (RLHF) \cite{ouyang2022training}, aligns model outputs with human preferences but is inefficient. Direct Preference Optimization (DPO) \cite{rafailov2023direct} and its variants \cite{azar2024general, chowdhuryprovably} address this by directly adjusting model policies using preference data. However, they face challenges like sample sparsity in recommendation and advertising. Group Relative Preference Optimization (GRPO) \cite{shao2024deepseekmath} uses explicit reward functions and group-based evaluation to enhance decision-making and training efficiency. In this paper, we present the first exploration of applying the generative model to bid shading, leveraging reward systems to guide the model beyond the constraints of traditional methods.

\section{Preliminaries}

\subsection{Problem Definition}
Suppose each DSP has an online advertising system to process bidding requests in the RTB scenario. When users visit web pages containing advertising opportunities, each DSP selects the candidate advertisement with the highest advertising value and places a bid. Once the auction ends, the winning advertisement is displayed, and the win/loss tags are returned to the DSP. Let  $X = \left\{x_{1}, x_{2},\ldots, x_{N}\right\}$ represent the bid requests, where $x_i$  includes features such as user and contextual information. The value of each request is calculated through metrics such as the click-through rate (CTR) and conversion rate (CVR), denoted as $v = \left\{v_{1}, v_{2},\ldots, v_{N}\right\}$, which signifies the expected value derived from an advertisement request and serves as the pre-shadow bid$\footnote[1]{This step usually solves an optimization problem with constraints and will not be elaborated on in detail in this paper.}$. In the bid shading system, we determine a shading ratio $\alpha_i$  for each request using a generative model, and multiply it with the unshaded bid to obtain the final bid $B = \left\{v_{1} * \alpha_1, v_{2} * \alpha_2,\ldots, v_{N} * \alpha_N\right\}$. The expected surplus of a bid request $x_{i}$ at the bid price $b_{i}$  is $surplus(x_i,b_i)=(v_i-c_i) \cdot Wr(b_{i} \mid x_{i})$, where $Wr(\cdot)$ represents the winning rate and $c_i$ represents the cost incurred if this request is won, $c_i = b_i$  in FPA scenario. For the bid shading system, the bid prices $b^{*}_i$ maximize the expected surplus:
\begin{equation}
b^{*}_i =
 \underset{b_i\in(b_{min}, b_{max})}{\arg\max }(v_i-c_i) \cdot Wr(b_{i} \mid x_{i}),
\end{equation}
where $b_{min}$ and $b_{max}$ are manually set as the lower and upper limits for bid shading. The DSP initiates a bid using the final $b^{*}_i$  and compares it against the bids from other advertisers. Once the auction concludes, the win/lose label $y^{*}_i$  and the final cost  $c_i$  are determined. Suppose $z_i$ represents the winning price, and we have
\begin{equation}
y_{i}^{*} = 
\begin{cases} 
1, & b_{i}^{*} > z_{i}, \\ 
0, & b_{i}^{*} \leq z_{i}. 
\end{cases}
\end{equation}

It is important to note that $z_i$ is not disclosed to the DSP. The actual surplus of the bid request $x_i$ at $b_i^*$ is ${surplus}(x_i, b_i^*) = (v_i - c_i) \cdot y_i^*,$  which serves as a crucial evaluation metric for bid shading.

\subsection{Generative Model Preprocessing}
Many methods output a scalar by discretizing continuous values and converting the prediction problem into a classification task to simplify learning. \cite{sun2024cread, lin2023tree}. However, predictions for the discretized buckets are typically generated independently, limiting opportunities for effective error correction. Current generative recommendation methods frequently utilize residual quantization (RQ-VAE) and hierarchical K-means clustering to obtain index IDs, and generate ID sequences through an autoregressive model \cite{rajput2023recommender}. In addition, generative regression \cite{ma2024generative} also uses this way to model scalar generation, achieving excellent performance. Inspired by these methods, we propose generating the shading ratio using an autoregressive encoder-decoder architecture.

Specifically, we introduce a vocabulary $\mathcal{V}=\left\{ w_j \right\}_{j=1}^{V}
$  where each element $w_i$  represents a predefined scalar value, analogous to tokens in the language model. The vocabulary embedding matrix is denoted as ${E} \in \mathbb{R}^{V \times D}$, where $V$ consists of the vocabulary size, and $D$ is the dimension of the token embeddings. Each shading ratio can be encoded as a sequence of tokens $s_{i} = \left\{ s_{i}^{1}, \ldots, s_{i}^{L} \right\}$, with $L$ indicating the length of the sequence. This process is known as tokenization. In contrast to generative recommendation systems that index directly based on ID, we propose a label decoding function  $g (\cdot)$  to determine the actual scalar value from the token ID output by the model, thereby deriving the shading ratio $\alpha_i = g(s_i) = \sum_{l=1}^{L} g(s_i^l) \in \mathbb{R} $.

\begin{figure*}[!t]
\centering
\includegraphics[scale=0.35]{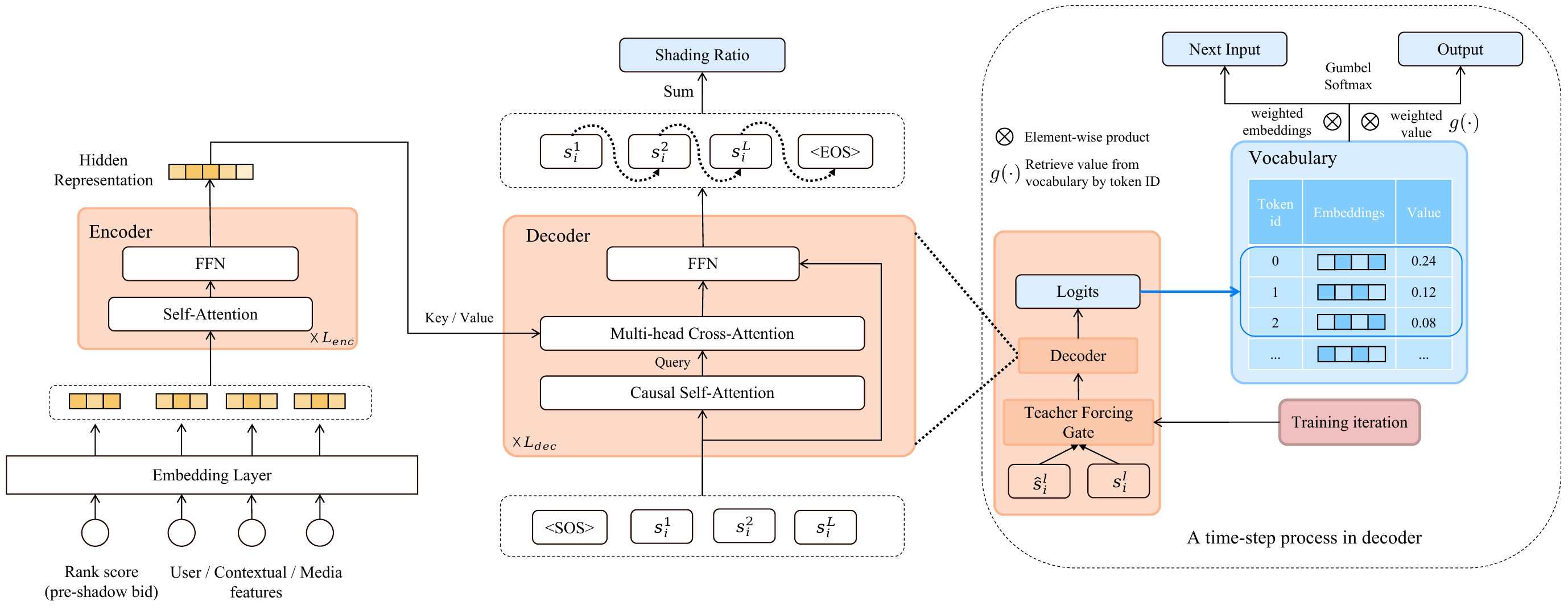}
\caption{The proposed generative model employs an encoder-decoder architecture to predict the token sequence autoregressively. It determines the specific value of each token using a decoding function $g(\cdot)$, then accumulates these values to obtain the final shading ratio. Additionally, it utilizes the gumbel softmax and teacher forcing strategies to ensure stability and accelerate convergence.}
\label{model}
\end{figure*}

\section{Method}
In this section, we provide a detailed explanation of the proposed GBS. First, we introduce the end-to-end autoregressive generative model. Next, we present the reward preference alignment system, which employs a channel-aware hierarchical dynamic network as the reward model and utilizes GRPO to guide policy learning via two reward alignment modules.

\subsection{End-to-End Generative Model}
\label{section4.1}
Our generative model decomposes the final task into multiple classification subtasks, utilizing inter-step dependencies to approximate the final shading ratio accurately.  This framework allows each prediction step the flexibility to choose from the vocabulary, and the expanded solution space enables the model to generate a broader spectrum of potential sequences.

\textbf{Tokenizer}. The vocabulary $\mathcal{V}$ comprises predefined ratio slots as tokens, allowing the model to generate shading ratios that closely approximate actual values. Two key properties are prioritized when constructing $\mathcal{V}$: 1) Each token must be unique, enabling $\mathcal{V}$ to represent all possible shading ratios with a finite token set;   2) Token frequencies should be relatively uniform to prevent class imbalance. We employ the dynamic percentile-based algorithm \cite{ma2024generative} for automated vocabulary construction from specified datasets. It accelerates tail-value reduction, rapidly decreases inter-update variance, and enhances algorithmic reliability (see Appendix \ref{appendixA} for details). 

Due to the lack of ground-truth labels for shading ratios, we construct the vocabulary $\mathcal{V}$ using the shading ratio from the two-stage method and perform tokenization, which involves transforming ratios $\left\{\alpha_{i}\right\}_{i=1}^{N} 
$ into discrete tokens $\{ s_{i} = \{ s_{i}^{1}, \ldots, s_{i}^{L} \} \}_{i=1}^{N} 
$. To reduce the complexity of learning encoded labels, the tokenized sequences should adhere to three principles: 1) The reconstructed value $g(s_i)$ should minimize error relative to the real ratio;
 2) The sequence length $L$ of $s_i$ should be minimized;
 3) The sequence should maintain monotonicity: $s_{i}^{1} \geq s_{i}^{2} \geq \ldots \geq s_{i}^{L}$.
Sequences that meet these conditions enable the model to incrementally output residual results, thereby significantly simplifying the learning process. Consequently, we employ a greedy strategy during tokenization, generating training data for the pre-training stage.

\textbf{Encoder-Decoder Architecture}.
As illustrated in Fig. \ref{model}, our model employs a Transformer-based architecture, consisting of an embedding layer, an encoder for modeling features, and a decoder for generating shading ratios. Specifically, the embedding layer maps features through the parameter matrix. For dense features, we transform them into the same dimension using an affine transformation:
\begin{equation}
{e}_{i j}\in \mathbb{R}^{1 \times d}= \begin{cases}{W}_j * x_{i j}+{b}_j, & x_{i j} \text { is a continuous feature,} \\ \operatorname{lookup}\left({E}_j, e_{i j}\right), & x_{i j} \text { is a sparse feature,}\end{cases}
\end{equation} 
where $W$ and $b$ are learnable parameters. Then, we concatenate them to get the initial embedding representation  of $x_i$: 
\begin{equation}
{e}_i \in \mathbb{R}^{f \times d}=\left[{e}_{i1} , {e}_{i2}  , \ldots , {e}_{if} \right] ,
\end{equation}
where $d$ is the dimension size of each field, and $f$ is the number of feature fields. The encoder processes input features using stacked multi-head self-attention and feed-forward layers:
\begin{equation}
H_i = (...FFN({SelfAttn}(e_i))). 
\end{equation}

To standardize decoder training, we introduce the target sequence with a start-of-sequence token $<SOS>$ and padding tokens $<PAD>$. These tokens do not represent any semantic value in the label space  (i.e., $g(\{<SOS>, <PAD>\})=0$). The decoder autoregressively generates the target sequence based on the encoder’s output $H_i$ and the preceding subsequence. At training timestep $l$, the output token is
\begin{equation}
\label{argmax}
\hat{s}_{i}^{l} = \arg\max_{w \in \mathcal{V}} P_{\theta}\left(w \mid {H}_{i}, \hat{s}_{i}^{<l}\right).
\end{equation}

\textbf{Pre-training Optimization}.
During pre-training, we equip the model with fundamental market perception capabilities through supervised fine-tuning (SFT):

1) Cross-entropy loss for next-token prediction:
\begin{equation}
{L}_{NTP} = -\sum_{i=1}^{N} \sum_{l=1}^{L} \log P_{\theta}\left(\hat{s}_{i}^{l} \mid {H}_{i}, \hat{s}_{i}^{<l}\right),
\end{equation} where labels are derived from shading ratios generated by a two-stage method, we adopt a teacher forcing (TF) strategy to improve model efficiency, which directly feeds the ground truth $s_i^l$ as input at step $l+1$  to guide model training. We design a gate to enable the TF strategy adaptively:
\begin{equation} TF_{pro} = TF_{pro} \cdot \frac{\eta}{\eta + e^{\left( \frac{iteration}{\eta} \right)}},
\end{equation}
where $TF_{pro}$ denotes the probability of employing the TF strategy, and $\eta$ is a hyperparameter. The gate mechanism decreases the probability of using TF as the number of training iterations increases, promoting faster convergence and improved model stability.

2) Negative Log-Loss of expected surplus for winning data $N_{+}$:
\begin{equation}
L_{\text{surplus}} = -\sum_{i=1}^{N_{+}}\mathbb{E}\left( \log\left( (v_i - C(v_i \cdot \alpha_i)) \cdot Wr(v_i \cdot \alpha_i \mid x_i) \right) \right),
\end{equation}
where $\alpha_i$ is the predicted shading ratio, $C(\cdot)$ is the cost function ($C(b)=b$ in FPA), and $Wr(\cdot)$ is the estimated winning rate (see Section \ref{section4.2.1}). Notably, the argmax operation in Equation \ref{argmax} obstructs gradient backpropagation. Therefore, we employ the gumbel softmax to obtain a differentiable distribution during token generation. 
At time step $l$, the dimension of the estimated raw output vector is $V$ (the size of the vocabulary). We do not consider the entire vocabulary space when calculating the weighted vector. Since the vocabulary is strictly ordered, we focus on the domain $V'$, which corresponds to the neighborhood of the indices of the highest output values. For each position $p \in V'$, we sample $u_p$ from the uniform distribution $U \sim \operatorname{Uniform}(0,1) $, and generate gumbel noise $n_{p} = -\log\left(-\log\left(u_{p}\right)\right) $ through the inverse transform, adding it to the original logits:

\begin{equation}
weight_{p}^l=\frac{\exp\left(\left(l_{p}+n_{p}\right)/\tau\right)}{\sum_{j=1}^{V'}\exp\left(\left(l_{p}+n_{p}\right)/\tau\right)},
\end{equation} where $l_{p}$ is the logit predicted by the decoder at step $l$ and position $p$, and $\tau$ is the temperature coefficient. As $\tau$ approaches 0, the output resembles discrete sampling; as $\tau$ approaches infinity, the output approximates a uniform distribution. Finally, we modify the decoding function $g(\cdot)$ using gumbel softmax weights and use the embedding of weighted sum $\hat{s}_{i}^{l}$ as input for the next token prediction. Compared to argmax, gumbel softmax not only addresses the issue of differentiability but also reduces the impact of errors in the earlier positions of the output sequence, increasing the stability.

In the pre-training stage, the loss is  ${L} = {L}_{NTP} + \lambda \cdot {L}_{surplus}$, where $\lambda$ is a hyperparameter.

\subsection{Reward Preference Alignment System}
% Pre-trained models fit existing two-stage solutions, thereby confining their ability to exceed the performance limits of conventional approaches.    Although the negative logarithm of expected surplus steers the model toward surplus extrema, the non-unimodal nature of the surplus loss curve results in challenges similar to those encountered with two-stage solutions, susceptibility to local optima.  Furthermore, training the loss function only on winning bids introduces significant bias.    To overcome these issues, we implement reward-based preference alignment during post-training.    Employing policy reinforcement learning, the model is trained in a generative item space, where reward signals facilitate the perception of finer-grained preference information and help the model develop a global perspective to escape local optima.    Specifically, a surplus preference reward discourages overbidding on acquirable bid requests, while an exploratory utility reward balances the information volume against costs, thus mitigating data selection bias in bidding scenarios.
Pre-trained models are limited by two-stage solutions and often get stuck in local optima due to the SFT in non-convex surplus loss and bias from training only on winning bids.   To address this, we use reward-based preference alignment during post-training.   With policy reinforcement learning, the model receives reward signals that help it develop a global perspective to escape local optima.   Specifically, a surplus preference reward prevents overbidding, while an exploratory utility reward balances information and cost. In this section, we first introduce the reward model and then introduce the two reward preference alignment systems.

\begin{figure}[!t]
\centering
\includegraphics[scale=0.33]{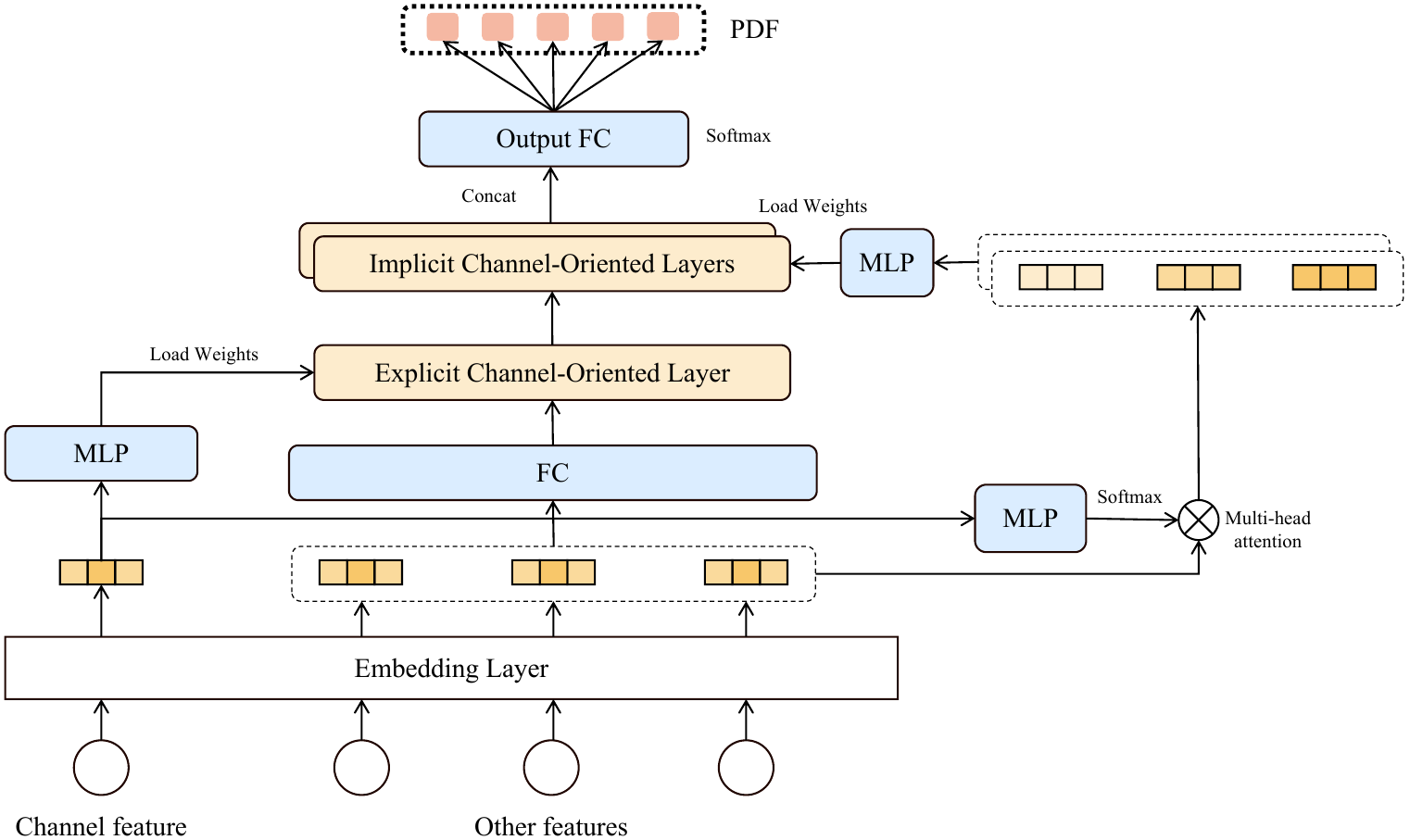}
\caption{The architecture of CHNet.}
\label{CHNet}
% \vspace{-0.5cm}
\end{figure}

\subsubsection{Channel-Aware Hierarchical Dynamic Network (CHNet)}
\label{section4.2.1}

Bid landscape forecasting is critical for evaluating bid quality. Following the previous works \cite{ren2019deep, huang2024second}, we first discretize the continuous space via equal frequency binning, where a set of $T$ prices $0 < b_{1} < b_{2} < \ldots < b_{T} $ covers the finite precision of price determinations. We have the probability density function (PDF) $P(z=b \mid x)$ of the winning price $z$ being $b$, and the cumulative distribution function (CDF) $ Wr(b_{t} \mid x) = P(z\leq b_{t}\mid x)=\sum_{j\leq t} P(z\in B_{j})$ representing the winning rate of a certain bid $b_t$, where $t = 1, 2, ...T$ and $B_{j}=(b_{j}, b_{(j+1)}]$ is a divided disjoint interval. Once the PDF curve of a request bid is obtained, the winning rate for another bid can be derived with a single additional CDF integral calculation.

In real-world advertising systems, varying media employ opaque and diverse billing methods, leading to significant differences in the bid landscape across channels. Inspired by multi-scenario approaches \cite{gao2023scenario}, we propose a channel-aware hierarchical dynamic network (CHNet) for fine-grained channel information modeling, as shown in Fig. \ref{CHNet}. CHNet utilizes a hierarchical structure with an explicit channel-oriented layer and multiple implicit channel-oriented layers, sequentially modeling coarse-grained explicit information and fine-grained implicit information. For processing feature matrices\footnote[2]{for simplicity, request $i$ is not separately denoted in this section.}, we adhere to the method outlined in Section \ref{section4.1}, with separate processing for channel feature $e^c$  and others $e^o$.

\textbf{Explicit channel-oriented layer.} We perform global feature interactions for the common feature $e^o$ using a multi-layer perceptron (MLP), denoted as $e^{global} = MLP(e^o)$. Furthermore, we adaptively generate the weight parameters based on the channel information $e^c$  via a re-parameterization approach, specifically by applying an MLP followed by a reshape operation: $W_{exp},b_{exp} = Reshape(MLP(e^c))$. These parameters are then used to instantiate the explicit scenario-oriented layer, resulting in the explicit channel representations:
\begin{equation}
e^{{explicit}} = {FC}(e^{{global}}; W_{exp},b_{exp}).
\end{equation}

\textbf{Implicit channel-aware module.} After explicit channel modeling, CHNet uncovers implicit patterns using a channel-aware multi-head attention mechanism. This enables comprehensive modeling of complex data distributions and adaptive identification of key implicit patterns. First, channel embeddings $e^c$ are processed by a MLP to generate raw weights, which are then normalized into $G$ groups ($G$ is the number of attention heads) using the softmax:

\begin{equation}
\begin{gathered}
{weight}_{{ori}} = {Reshape}\left({MLP}\left(e^{c}\right)\right), \\
{weight}_{{norm}}[g] = {Softmax}\left(\text{weight}_{{ori}}[g]\right),g \in [1, G].    
\end{gathered}
\end{equation}

Next, we perform an element-wise product of the normalized weight with the feature embedding: $e^{weight} = {weight}_{{norm}} \otimes {e}^{o}
$. 

Then, using the re-parameterization technique, we generate the weight parameters for different implicit channel layers, denoted as
\begin{equation}
{W}_{{imp}}[g], {b}_{{imp}}[g] = {Reshape}({MLP}({e^{weight}}[g])), g \in [1, G].
\end{equation}

We input the explicit representations into each implicit channel-oriented layer for implicit modeling. After that, 
we concatenate the outputs of all implicit layers to predict the PDF distribution:
% \begin{equation}
% {e}_{g}^{{implicit}}= {FC}(e^{{explicit}}; W_{imp}[g],b_{imp}[g]). 
% \end{equation}
\begin{equation}
\begin{gathered}
{e}_{g}^{{implicit}}= {FC}(e^{{explicit}}; W_{imp}[g],b_{imp}[g]), \\
output = {Softmax}({FC}({Concat}({e}_{1}^{{implicit}},\dots,{e}_{G}^{{implicit}}))).
\end{gathered}
\end{equation}

\textbf{Training loss.}
During the CHNet training process, the loss function is comprised of three components. For winning bid logs in a non-FPA scenario, we directly compute the negative log-likelihood loss based on the actual winning price: ${L}_{win} 
= -\sum_{i=1}^{N_{+}} \log p\left(c_{i} \mid x_{i}\right)$. Note that in the FPA scenario, this loss cannot be applied because the cost paid does not reflect the minimum winning price. For losing bid logs, we only know that the winning price exceeds the current bid, and we maximize the CDF of the right half of the bid, resulting in ${L}_{lose}
= -\sum_{i=1}^{N_{-}}  \log[1 - Wr(b_{i} \mid x_{i})]$. Additionally, to further enhance the model’s ranking and classification capabilities \cite{lin2024understanding}, we introduce a pairwise loss related to the winning rate, denoted as 
\begin{equation}
{L}_{{ranknet}}=-\frac{1}{N_+N_-}\sum_{i=1}^{N_+}\sum_{j=1}^{N_-}\log(\sigma(Wr(b_i|x_i)^{(+)}-Wr(b_j|x_j)^{(-)})).
\end{equation}

Finally, the loss function of CHNet is formulated as
\begin{equation}
{L} =\lambda_1 \cdot  {L}_{win} + \lambda_2 \cdot {L}_{lose}+ \lambda_3 \cdot {L}_{ranknet}.
\end{equation}

\begin{figure}[!t]
\centering
\includegraphics[scale=0.57]{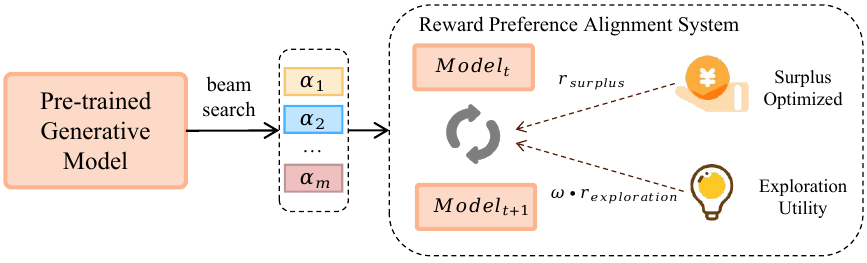}
\caption{The architecture of post-training with GRPO.}
\label{GRPO}
% \vspace{-0.5cm}
\end{figure}

\subsubsection{Surplus Optimized Alignment}
The primary objective of bid shading is to prevent incurring excessive costs. Accordingly, we use the expected surplus associated with the shaded bid $b=v*\alpha$ as the reward $r_{surplus} = (v-C(b)) \cdot Wr(b \mid x)$ and optimize the model using reinforcement learning strategies. We employ GRPO to optimize the strategy based on relative rewards within the group. More specifically, for each bid request $req$, GRPO samples a group of outputs $\{ \alpha_{1}, \ldots, \alpha_{2}, \ldots, \alpha_{m} \}$ from the old policy and then optimizes the policy model by maximizing the following objective:
\begin{equation}
\label{jGrpo}
\begin{aligned}
&\mathcal{J}_{\text{GRPO}}(\theta) = \mathbb{E}_{req\sim P(R), \{\alpha_i\}_{i=1}^m \sim \pi_{\theta_{\text{old}}}} 
\Bigg[ \frac{1}{m}\sum_{i=1}^m \min \bigg( \frac{\pi_\theta(\alpha_i \mid req)}{\pi_{\theta_{\text{old}}}'(\alpha_i \mid req)} A_i, 
\\ &\operatorname{clip}\Big(\frac{\pi_\theta(\alpha_i \mid req)}{\pi_{\theta_{\text{old}}}'(\alpha_i \mid req)}, 1-\epsilon, 1+\epsilon\Big) A_i \bigg)  - \beta \cdot \mathbb{D}_{\text{KL}} \big[ \pi_\theta \| \pi_{\text{ref}} \big]  \Bigg],
\end{aligned}
\end{equation}
where $A_i $ denotes the advantage, which is calculated solely based on the relative return of each group's internal output:
\begin{equation}
\label{Ai}
A_i = \frac{r_i - \operatorname{mean}(\{r_1,r_2,...,r_m\})}{\operatorname{std}(\{r_1,r_2,...,r_m\})}.
\end{equation}

Additionally, we add the KL divergence between the training and reference strategies in the objective. GRPO simultaneously employs the group-relative advantage estimation and a KL divergence penalty to ensure that strategy updates are efficient and stable. The complete algorithm processing is detailed in Algorithm \ref{post-training}.

Pre-training with surplus-based SFT restricts the model to exploring the specific output bid and may cause gradient descent to get stuck in local optima due to the non-convex surplus.   In contrast,  policy reinforcement learning expands the generation space for getting global optima. Since the bidding process is uncertain and each winning price reflects a probability distribution, we incorporate losing bid data with a smaller weight during post-training.

% Although pre-training employs surplus-based SFT, there are substantial differences between the two stages. During the SFT stage, the model can only explore the current bid point along the surplus curve. Gradient descent can become trapped in local optima due to non-convex surplus. In contrast, policy reinforcement learning adopts a broader perspective, enhancing the ability to identify global optima. Furthermore, unlike the SFT stage, which only optimizes within the winning data, lost requests are also important. The bidding process is inherently uncertain, as each winning price corresponds to a probability distribution rather than a fixed value. Therefore, we utilize losing bid data in the post-training stage with a smaller weight.

\subsubsection{Exploration Utility Alignment}
In the DSP advertising bidding scenario, data selection bias presents a significant challenge, as the collected data are generated using a fixed bidding algorithm, which complicates the bid landscape forecasting. To address this issue, we incorporate bid exploration into the bid shading system, providing more robust data support for maximizing long-term surplus. 

According to uncertainty sampling in active learning \cite{settles2008analysis, wang2024cost}, we think that the greater the predicted uncertainty, the more informative the exploration will be. We quantify the exploration utility using the entropy of the winning rate. For a bid request $x_i$, the predicted winning rate $Wr_i= Wr(b_i \mid x_i) $ at a particular bid $b_i$ can be framed as a binary classification problem (win or lose). Thus, the entropy of the bid request at bid $b_i$ is denoted as 
 \begin{equation}
\phi^E(x_i,b_i)=-( Wr_i \cdot log( Wr_i )+(1- Wr_i ) \cdot log(1- Wr_i)).
 \end{equation}
 
Additionally, during exploration, it is important to avoid excessive cost increases; the final exploration reward is denoted as $r_{exploration} =\frac {\phi^E(x_i,b_i)}{C(b_i)}$. When calculating the final advantage $A_i$ (Equation \ref{Ai}), the reward is 
\begin{equation}
\label{final_r}
r = r_{surplus} + \omega \cdot r_{exploration},
\end{equation}
where $\omega$ is not a fixed parameter. It is unnecessary to explore all samples; by identifying samples with high similarity, we can avoid unproductive exploration of outliers. Inspired by the metric learning \cite{snell2017prototypical}, we compute the exploration weight for request $x_i$:
\begin{equation}
\omega_i = cos(e_i,\frac{1}{\left|N\right|}\sum_{j=1}^{N}e_j),
\end{equation}
where cos(·, ·) is the cosine similarity function, more representative requests are assigned greater exploration weight. The post-training structure is shown in Figure \ref{GRPO}.

\section{Experiments}
% where implementation details are detailed in Appendix \ref{Implementation}.
\subsection{Experimental Setup}

\subsubsection{Datasets}
We evaluate our method on both the public and real-world industrial datasets, with detailed statistics presented in Table \ref{datasets}.    The public iPinYou \cite{liao2014ipinyou} data are derived from the second-price auctions setting, where the advertiser’s bid is used as the actual value in the bid request, and the paid price is regarded as the winning price.    The dataset is randomly split into training and testing sets with a ratio of 7/3.   Our private dataset is collected from one month of bid sampling data from Meituan DSP and is partitioned similarly.    Notably, only the testing set contains minimum  winning price, which is used for evaluation.

\begin{table}[htb]
\caption{The statistics of datasets.}
\label{datasets}
\begin{tabular}{cccc}
\toprule
\textbf{DataSet} & \textbf{Size} & \textbf{Winning rate} & \textbf{features} \\ \hline
iPinYou          & 10,577,061    & 29.7\%                & 18                \\
Private          & 162,549,577   & 3.89\%                & 197               
\\ 
\bottomrule
\end{tabular}
\end{table}
% \vspace{-5mm}

\subsubsection{Evaluation Metrics}
For offline evaluation in FPA, business performance is crucial in determining whether an algorithm is suitable for production deployment, as it reflects the algorithm’s effectiveness. Given the objective of bid shading, we use surplus $(V-b) \cdot I(b>z)$ and surplus rate as key business metrics, as they directly indicate the impact of the algorithm on business outcomes. Specifically, we calculate the surplus rate as the percentage of residual surplus relative to the total optimal surplus:

\begin{equation}
\text{surplus rate (SR)} = \frac{\sum_{i}(v_{i} - b_{i})I(b_{i} > {z}_{i})}{\sum_{i}(v_{i} - z_{i})}.
\end{equation}

\subsubsection{Implementation Details}
We implement all deep learning baselines and the proposed GBS on NVIDIA A100-SXM4-80GB GPUs. During the pre-training stage, we use the Adam optimizer with a batch size of $2^{10}$ and a learning rate of $1e^{-4}$. The maximum output sequence length $L$ is set to 5, $\eta$ in the TF Gate is 5000, and the temperature coefficient $\tau$ in gumbel softmax is 0.2. In the post-training stage, the loss hyperparameters $\lambda_1$, $\lambda_2$, and $\lambda_3$ for CHNet training are set to 0.5, 1, and 0.2, respectively, while the KL weight $\beta$ in the GRPO optimization function is set to 0.01. We divide the training and testing sets of the iPinYou and private datasets into 100 segments each to simulate the continuous arrival of bid requests in an online environment.  We calculate the surplus by the winning price in each round and collect the win/loss labels.  Subsequently, the bid data are used to update the CHNet and the generative model.  This process is repeated for 100 rounds until the experiment is complete.

\subsection{Overall Performance}
We compare the performance of GBS with several state-of-the-art baselines, including the bid shading and the generative methods:

\textbf{EDDN \cite{zhou2021efficient}:} A two-stage bid shading method, which optimizes the surplus using golden section search in the OR stage.

\textbf{WR \cite{pan2020bid}:} A two-stage bid shading method, which optimizes the surplus using bisection algorithm in the OR stage.

\textbf{TSBS-DLF \cite{ren2019deep}:} A two-stage bid shading method that uses the DLF model in the ML stage, which models the bid landscape without distributional assumptions,  employing the conditional probability chain rule and LSTM.

\textbf{TSBS-ADM \cite{li2022arbitrary}:} A two-stage bid shading method uses the ADM model in the ML stage, which uses neighborhood likelihood loss for accurate prediction.

\textbf{MEBS \cite{gong2023mebs}:} An end-to-end bid shading method, which jointly optimizes the shading model and the win rate model, performs supervised learning through the negative logarithm of the surplus as the loss.

\textbf{CVAE \cite{sohn2015learning}:} An extended model of variational autoencoder, which achieves sample generation based on specific inputs by integrating conditional variables in the encoder and decoder. We optimize it through the pre-training method proposed in this paper.

\textbf{DF \cite{ho2020denoising}:} A generative diffusion model that corrupts the data distribution by gradually adding noise and then learns the inverse denoising process to generate shading ratio. We optimize through the pre-training method proposed in this paper.

\textbf{Post-CVAE \& Post-DF:} Post-training will be performed on the pre-trained generative model, which uses reparameterization to transfer gradients.

Based on the results in Table \ref{Overall}, we can draw the following conclusions: 1) The traditional two-stage algorithm performs significantly worse on the Meituan private dataset, which suggests that a more complex market environment poses greater challenges for bid shading, and that coarse-grained bid landscape forecasting and bisection search cannot adapt to a broader range of business scenarios; 2) The traditional generative models show varying degrees of improvement after post-training and outperforms the two-stage method, indicating that policy reinforcement learning can effectively enhance these models to raise their performance ceiling, and the end-to-end generative approach offers distinct advantages; 3) GBS achieves outstanding results across all metrics.    By combining an autoregressive residual generative model with post-training, GBS effectively mitigates the suboptimality of traditional methods and adapts well to complex market environments.

\begin{table}[h]
\caption{Overall offline experiment results.}
\label{Overall}
\begin{tabular}{l|cc|cc}
\toprule
\multirow{2}{*}{Model} & \multicolumn{2}{c|}{\textbf{iPinYou}} & \multicolumn{2}{c}{\textbf{Private}}  \\
 & SR & Surplus    & SR & Surplus  \\
\midrule
EDDN                   & 47.82\%                & 90,199,768                   & 19.88\%                & 14,582,074                  \\
WR                     & 54.90\%                & 103,560,851                  & 26.13\%                & 19,164,882                  \\
TSBS-DLF               & 55.22\%                & 104,164,484                  & 27.25\%                & 19,986,132                  \\
TSBS-ADM               & 55.49\%                & 104,673,800                  & 32.67\%                & 24,183,220                  \\
MEBS                   & 54.46\%                & 102,716,112                  & 31.45\%                & 23,041,120                  \\ \midrule
CVAE                   & 51.42\%                & 96,981,112                   & 36.45\%                & 26,698,064                  \\
Post-CVAE              & \underline{57.14}\%                & \underline{107,781,128}                  & \underline{36.63}\%                & \underline{26,830,032}                  \\
DF                     & 48.04\%                & 90,617,280                   & 31.28\%                & 22,912,038                  \\
Post-DF                & 51.13\%                & 96,445,909                   & 33.89\%                & 24,823,816                  \\ \midrule
\textbf{GBS}           & \textbf{60.48\%} & \textbf{114,068,392} & \textbf{41.74\%} & \textbf{30,580,080} \\ 
\bottomrule
\end{tabular}
\end{table}
% \vspace{-5mm}

\subsection{Ablation Study}
To assess the effectiveness of each component in GBS, we conducted a series of ablation studies. Specifically, we build several variants of the GBS:

\textbf{w/o PostT:} A variant of GBS without the Post-training stage, which directly uses the pre-trained model to output shading ratio.

\textbf{w/o PostT-EUA:} A variant of GBS without the exploration utility alignment in post-training, which does not perform bid exploration and only performs surplus optimization.

\textbf{w/o GM:} A variant of GBS without the autoregressive generative model, which uses the traditional generative model CVAE \cite{sohn2015learning} to generate the shading ratio and keeps the pre-training and post-training consistent.

\textbf{w/o CHNet:} A variant of GBS without the CHNet, which uses a binary classification model to provide the predicted winning rate.

Table \ref{ablation} presents the results of the ablation study. CVAE is less effective than our GM, underscoring the importance of progressively outputting results through residuals in refined bid generation.   Excluding CHNet and bid exploration also leads to similar performance drops, indicating that a more fine-grained reward model and a broader range of training data are crucial for enhancing model performance.   Omitting post-training results in the most significant performance degradation, highlighting the limitations of relying solely on supervised learning for the bid shading task. The results suggest that policy reinforcement learning plays a vital role in helping the model overcome performance bottlenecks.

\begin{table}[h]
\caption{The contributions of different components of GBS.}
\label{ablation}
\begin{tabular}{l|cc|cc}
\toprule
\multirow{2}{*}{Model} & \multicolumn{2}{c|}{\textbf{iPinYou}} & \multicolumn{2}{c}{\textbf{Private}}  \\
 & SR & Surplus    & SR & Surplus  \\
\midrule
w/o PostT                                   & 56.22\%     & 106,036,888    & 36.91\%          & 27,041,464          \\
w/o PostT-EUA                               & \underline{60.16}\%     & \underline{113,458,768}    & \underline{41.54}\%          & \underline{30,429,570}          \\
w/o GM                                      & 57.14\%     & 107,781,128    & 36.63\%          & 26,830,032          \\
w/o CHNet                                   &       58.22\%      &        109,805,915       & 38.16\%          & 27,953,616          \\
\textbf{GBS}                                & \textbf{60.48\%}   & \textbf{114,068,392}      & \textbf{41.74\%} & \textbf{30,580,080} \\ 
\bottomrule
\end{tabular}
\end{table}
% \vspace{-5mm}

\subsection{Performance Analysis of CHNet}
In this section, we evaluate the performance of the proposed CHNet in bid landscape forecasting on private datasets. We use the area under the curve (AUC)  and binary cross-entropy (BCE) of winning rate as evaluation metrics. In the experiments, we compare the following competitive bid landscape forecasting methods: 1). Censorship Linear Model (CLM) uses the normal distribution as the prior for single-point estimation of the winning rate; 2). Deep Landscape Forecasting (DLF) model \cite{ren2019deep}; 3). Arbitrary Distribution Model (ADM) \cite{li2022arbitrary}; 4). CHNet-w/o EUA is the ablation experiment without the exploration utility alignment module during the post-training phase.

\begin{figure}[htb]
\centering
\includegraphics[scale=0.28]{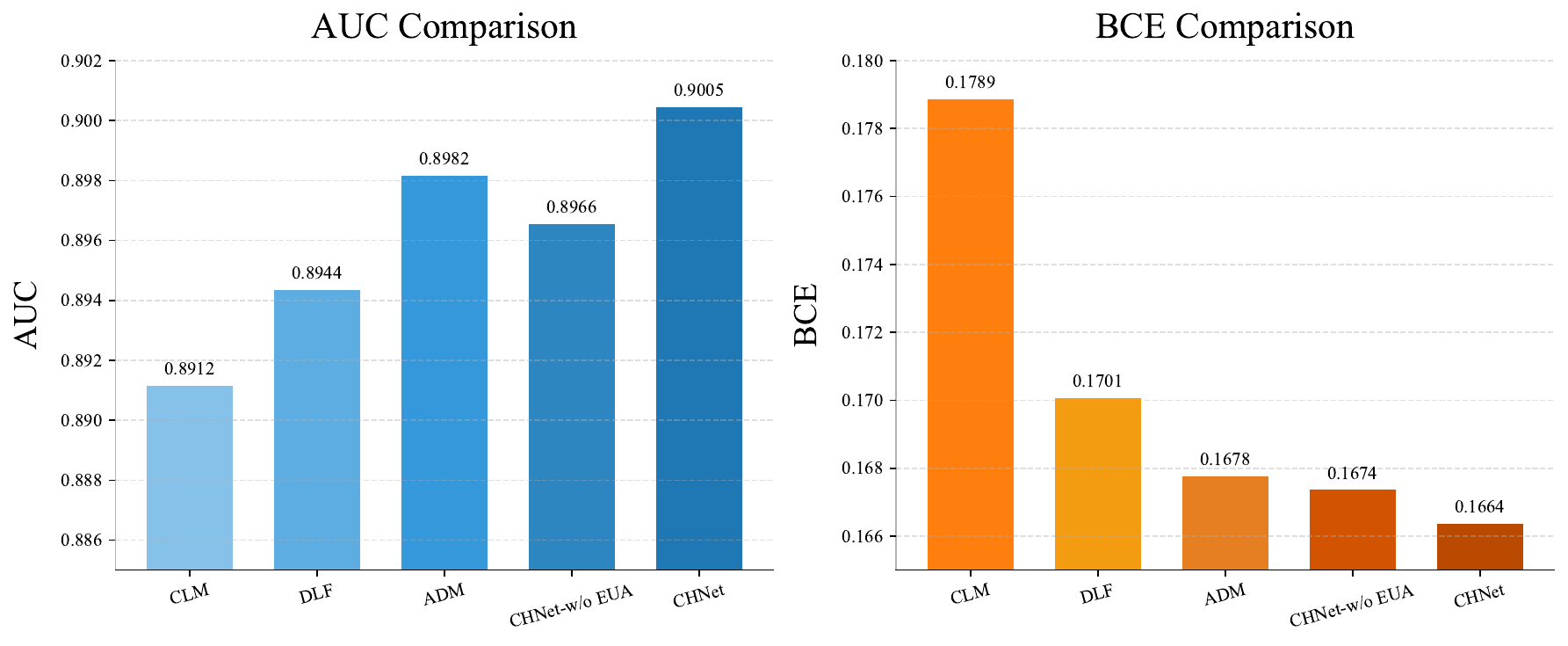}
\caption{The performance of CHNet.}
\label{CHNetexp}
\end{figure}
% \vspace{-5mm}
The experimental results are shown in Fig. \ref{CHNetexp}. The results demonstrate that CHNet performs superior ranking and classification tasks, owing to its refined modeling of channel information, a key characteristic of the market environment. Additionally, removing the bid exploration module decreases performance and exposes the model to data selection bias. Therefore, incorporating exploration in the bidding process is essential for improving the model’s generalization ability.

\begin{figure}[]
\centering
\includegraphics[scale=0.57]{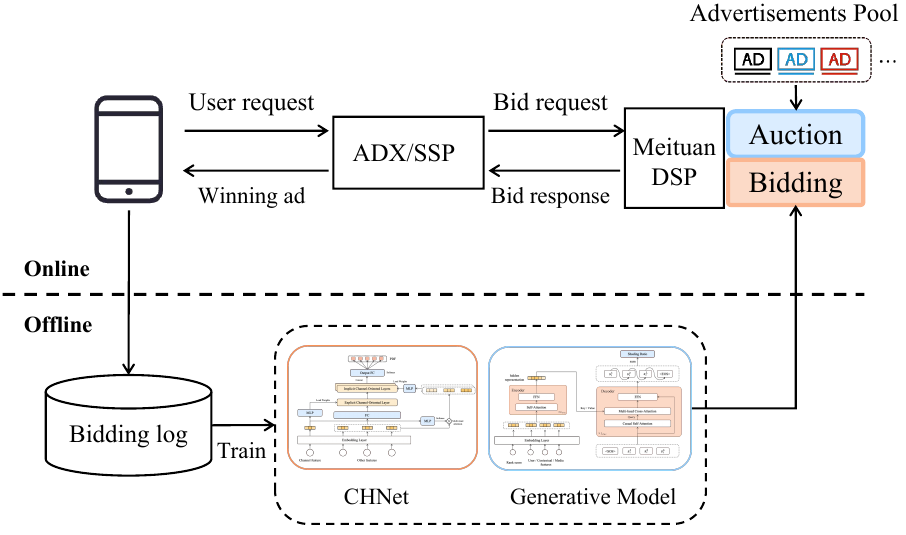}
\caption{Architecture of the online deployment with GBS.}
\label{MTOnline}
\end{figure}

\subsection{Performance on Online System (Meituan)}
To evaluate the online performance of GBS, we deployed it on the Meituan DSP. We conducted a rigorous A/B test over two weeks, and the engineering architecture is shown in Fig. \ref{MTOnline}.   We allocated 30\% of the traffic to GBS, while the baseline method was an online two-stage bid shading approach, which utilizes the golden section search in the operations research stage. We model certain channels as a non-ideal second-price auction scenario, consistent between baseline and GBS. We use four key online metrics for evaluation: Return On Investment (ROI), Cost Per Mille (CPM),  Cost Per Click (CPC), and Inference Time (IT).

Table \ref{A/B} shows the improvement of GBS relative to the baseline. The results indicate that GBS has significantly improved in all online metrics compared to the baseline, achieving significant business benefits. During the inference stage, GBS directly outputs the shading ratio, resulting in lower inference time than the two-stage method and reducing the burden on online services.

\begin{table}[htb]
\caption{Online A/B test results.}
\label{A/B}
\renewcommand{\arraystretch}{1.2}
\small
% \resizebox{0.45\textwidth}{!}{
\begin{tabular}{ccccc}
\toprule
                  & \textbf{ROI}   & \textbf{CPM}   & \textbf{CPC} &\textbf{IT}   \\ \hline
Baseline (Two-stage)          & 0.0\% & 0.0\% & 0.0\% & 0.0\%\\
\textbf{GBS}              & \textbf{+3.4}\%     & \textbf{-4.1}\%     & \textbf{-4.9}\%     & \textbf{-37.6}\%
\\ 
\bottomrule
\end{tabular}
% }
% \vspace{-0.3cm} 
\end{table}

\section{Conclusion and Future Work}
% In this paper, we propose a novel Generative Bid Shading (GBS) framework to address the suboptimality of existing bid optimization techniques.  GBS decomposes the bidding ratio into conditionally dependent sequences through an autoregressive generative model.  By a wide range of vocabulary space combinations with gumbel softmax and curriculum learning strategies, the model can accommodate complex distributions and accelerate convergence, thereby significantly enhancing the robustness of the generated results. GBS also introduces the CHNet reward model, which employs a hierarchical dynamic network to capture fine-grained characteristics of the bid landscape. Based on that, GBS incorporates dual modules for surplus optimization and exploration-utility alignment to prevent overbidding and mitigate sample selection bias, and utilizes the GRPO algorithm to enable efficient policy updates. Extensive offline and online experiments demonstrate that GBS outperforms state-of-the-art models. 
This paper proposes a novel Generative Bid Shading (GBS) framework that addresses the suboptimality of existing bid shading techniques.   GBS uses an autoregressive generative model to decompose the shading ratio into conditionally dependent sequences and incorporates a channel-aware reward model to capture fine-grained bid landscape features.   By integrating the reward preference alignment system with the GRPO, GBS prevents overbidding, mitigates sample selection bias, and enables efficient policy updates.   Extensive experiments show that GBS outperforms state-of-the-art models.

In the future, we will further optimize the generative model to better adapt upstream ranking scores and explore additional reward systems with advanced optimization algorithms.

% \bibliographystyle{ACM-Reference-Format}
% \bibliography{GBS}


% \newpage

\appendix

\section{Post-training process of GBS}
The post-training process is detailed in Algorithm \ref{post-training}.

\begin{algorithm}[hbp]
\caption{Post-training Stage: GRPO Optimization}
\label{post-training}

\KwRequire{    Pre-trained  generative model $\pi_{\theta_{\text{init}}}$,  reward system, 
    post-training data $\mathcal{D}$}
    
    Initialize policy model $\pi_{\theta} \leftarrow \pi_{\theta_{\text{init}}}$\;
\For{iteration $= 1$ \KwTo $I$}{
    Update reference model $\pi_{\text{ref}} \leftarrow \pi_{\theta}$\;
    \For{step $= 1$ \KwTo $S$}{
        Sample batch $\mathcal{D}_{b} \gets \mathcal{D}$\;
        Set old policy $\pi_{\theta_{\text{old}}} \gets \pi_{\theta}$\;
        \ForEach{bid request $req \in \mathcal{D}_{b}$}{
            Sample $m$ outputs $\{\alpha_i\}_{i=1}^m \sim \pi_{\theta_{\text{old}}}(\cdot\mid req)$\;
            Compute final rewards $\{r_i\}_{i=1}^m$  by Equation \ref{final_r}\;
            Estimate token advantages $\hat{A}_{i,t}$ via group-relative method (Equation \ref{Ai})\;
            Update $\pi_{\theta}$ by maximizing GRPO objective (Equation \ref{jGrpo})\;
        }
    }
}
    \KwResult{optimized policy model $\pi_{\theta}$}
\end{algorithm}
% \section{Implementation Details}
% \label{Implementation}

% \normalem
% \bibliographystyle{ACM-Reference-Format}
% \balance
% \bibliography{GBS}

\section{Vocabulary Construction}
\label{appendixA}
We use the shading ratios generated by the two-stage method as the base dataset to construct the vocabulary. The process (Algorithm \ref{CVDP}) is initialized with a high starting quantile $q_{start}$, and then adaptively reduced by a decay rate $\delta$ until reaching the terminal quantile $q_{end}$.

\begin{algorithm}[hbp]
\caption{Constructing Vocabulary with Dynamic Percentiles}
\label{CVDP}

\KwRequire{Shading ratios from two stage method $Y=\{\alpha_{j}\}_{j=1}^{N}$, initially empty Vocabulary $V=\{\}$, start percentile $q_{{start}}$, end percentile $q_{{end}}$, decay rate  $\delta$ , minimal restoration error $\epsilon_1$ and $\epsilon_2$.}
    
    Sort $Y$ in descending order to obtain $\hat{Y}=\{\hat{\alpha}_{j}\}_{j=1}^{N}$\;
    Initialize iteration counter $i=1$, error metric $err=\infty$, current percentile $q=q_{{start}}$\;
    
    \While{$err > \epsilon_1$}{
        Compute the $q$-percentile $o_i$ of $\hat{Y}$\;
        
        \If{$o_i \leq \epsilon_2$}{
            \textbf{break}\;
        }
        
        Generate a new token $w_i$ which satisfy $o_i=g(w_i)$ and insert $w_i$ into vocabulary $V$\;
        
        Update $\hat{Y}$ using:
        \[
        \hat{y}_{j} = \begin{cases}
            \hat{y}_{j}, & \text{if}\ \hat{y}_{j} < o_i, \\
            \hat{y}_{j} - o_i, & \text{otherwise}
        \end{cases}
        \]
        
        Update the error metric: 
        \[
        err = \max \left\{\frac{\hat{y}_{j}}{y_{j}}\right\}_{j=1}^{N}
        \]
        
        Update percentile $q$ with decay rate $\delta$: 
        \[
        q = \max(q \cdot \delta, q_{{end}})
        \]
        
        Increase $i$: $i \gets i + 1$\;
    }
    
    \KwResult{Vocabulary $V$}
\end{algorithm}

% \vfill
% \newpage 

% \newpage 

% \vfill\eject
\normalem
\bibliographystyle{ACM-Reference-Format}
\balance
\bibliography{GBS}

\end{document}